\documentclass{midl} % Include author names

\usepackage{xcolor}
\usepackage{hyperref}
\usepackage{booktabs}
\usepackage{multirow}
\usepackage{wrapfig}
\usepackage{colortbl}
\usepackage{arydshln}

\newcommand{\eg}{\textit{e.g.}}

\usepackage{mwe} % to get dummy images

\jmlrpages{}
\jmlryear{2024}

\jmlrworkshop{Short Paper -- arXiv preprint} 

\title[Rethinking Perceptual Metrics for Medical Image Translation]{Rethinking Perceptual Metrics for Medical Image Translation}

\midlauthor{\Name{Nicholas Konz\nametag{$^{1}$}} \Email{nicholas.konz@duke.edu}\\
\Name{Yuwen Chen\nametag{$^{1}$}} \Email{yuwen.chen@duke.edu}\\
\Name{Hanxue Gu\nametag{$^{1}$}} \Email{hanxue.gu@duke.edu}\\
\Name{Haoyu Dong\nametag{$^{1}$}} \Email{haoyu.dong151@duke.edu}\\
\Name{Maciej A. Mazurowski\nametag{$^{1,2,3,4}$}} \Email{maciej.mazurowski@duke.edu}\\
\addr $^{1}$Dept. of Electrical and Computer Engineering,
\addr $^{2}$Dept. of Radiology, 
\addr $^{3}$Dept. of Computer Science, \\
\addr $^{4}$Dept. of Biostatistics \& Bioinformatics,
\addr Duke University, NC, USA
}

\begin{document}

\maketitle

\begin{abstract}
Modern medical image translation methods use generative models for tasks such as the conversion of CT images to MRI. Evaluating these methods typically relies on some chosen downstream task in the target domain, such as segmentation. On the other hand, task\textit{-agnostic} metrics are attractive, such as the network feature-based perceptual metrics (\eg, FID) that are common to image translation in general computer vision.
In this paper, we investigate evaluation metrics for medical image translation on two medical image translation tasks (GE breast MRI to Siemens breast MRI and lumbar spine MRI to CT), tested on various state-of-the-art translation methods.
We show that perceptual metrics do not generally correlate with segmentation metrics due to them extending poorly to the anatomical constraints of this sub-field, with FID being especially inconsistent. However, we find that the lesser-used \textit{pixel-level} SWD metric may be useful for subtle intra-modality translation. Our results demonstrate the need for further research into helpful metrics for medical image translation.
\end{abstract}


\section{Introduction}
Medical image translation is the task of altering a medical image to look like it was taken in a different scanner or modality, and it has requirements that differ from mainstream natural image translation/style transfer. Namely, it is crucial to preserve anatomical/semantic content during translation. While natural image translation papers often employ perceptual learned-feature metrics like FID \cite{fid}, these metrics are not suitable for medical images because they may fail to capture global anatomical consistency and realism, as demonstrated previously \cite{jayasumana2023rethinking,contourdiff,segdiff} and in this work. A common metric that is more anatomy-focused is the performance of a segmentation model trained in the target domain on translated images \cite{vorontsov2022towards,kang2023structure} (or the reverse \cite{yang2019unsupervised}). 
Yet, this not only requires annotations for training and evaluating the segmentation model, but it may also be overly reductionist, leading to biased evaluations for specific tasks. In reality, translated images could be used for many applications, urging us to investigate better usage, understanding, and future development of \textit{downstream task-agnostic} metrics, \eg, perceptual metrics.

\section{Methods}

We evaluate two medical image translation tasks: (1) subtle \textit{intra-}modality breast MRI translation and (2) a more drastic \textit{inter-}modality translation of lumbar spine MRI to CT.

\paragraph{Datasets and Preprocessing.}

We use the public DBC dataset \cite{saha2018machine} for our dataset of 2D pre-contrast breast MRI slice images, following the same subsetting, preprocessing, and splits of \cite{segdiff}, with accompanying breast and fibroglandular tissue (FGT) segmentations from \cite{lew2024publicly}. Each split contains scans from both Siemens-manufactured (source domain) and GE-manufactured scanners (target domain). For lumbar spine MRI-to-CT slice translation, we follow the same preprocessing and train/validation/test splits of \cite{contourdiff} which did MRI-to-CT translation, using the same private lumbar spine MRIs for the source domain, and lumbar spine CTs from the public TotalSegmentator dataset \cite{Wasserthal_2023} for the target domain, both of which are accompanied by bone segmentations. All images are resized to $256\times 256$ and normalized linearly to $[0,255]$. Altogether, this forms train/validation/test splits of data from the source and target domains of $\{4096, 7900\}/\{432,1978\}/\{688,1890\}$ images for breast MRI, respectively, and $\{495, 1466\}/\{175,409\}/\{158,458\}$ for MRI-to-CT.

\paragraph{Translation Models.}
We evaluate unpaired image translation models of CycleGAN \cite{cyclegan}, MaskGAN \cite{phan2023structure}, UNSB \cite{kim2024unpaired}, and SPADE \cite{park2019SPADE}.
Unlike the other models, SPADE is an anatomically-guided segmentation-conditional model (denoted as ``SPADE$^\dagger$''), trained on target domain images \textit{and segmentation} pairs to generate translations from source domain image segmentations. All models were trained on target and source domain images (or only target domain images for SPADE) from the training set and evaluated using source domain images from the test set.

\begin{wrapfigure}[17]{htbp}{0.5\textwidth}
\includegraphics[width=0.5\textwidth]{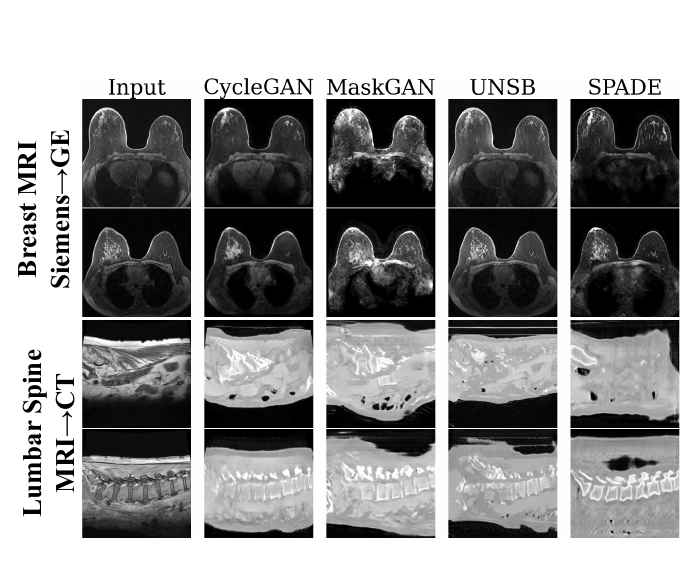}
\caption{Example translations for each model.}
\label{fig:imgs}
\end{wrapfigure}

% \subsection{Image Translation Evaluation Metrics}
% \label{sec:metrics}

% \begin{figure}[htbp]
% \centering
%  \includegraphics[width=0.8\textwidth]{figs/translation_results_v2.pdf}
% \caption{Example translations for each model.}
% \label{fig:imgs}
% \end{figure}

\paragraph{Segmentation Metrics for Image Translation.}
We first evaluate the metric of (1) training and validating a UNet on the corresponding target domain training set to segment the aforementioned object(s) of interest, and (2) evaluating it on test set images translated from the source domain with respect to their original segmentation labels, via the average prediction Dice coefficient.

\paragraph{Perceptual Metrics for Image Translation.}
We also assess perceptual metrics that compare translated image quality to real target domain images. These include FID \cite{fid}, the distance between Gaussian feature distributions extracted by an ImageNet-pretrained Inception network; KID \cite{binkowski2018demystifying}, like FID but without the Gaussian assumption and better for small datasets; and IS \cite{salimans2016improved}, which assesses feature diversity and quality. We also test the less common SWD \cite{karras2018progressive}, which focuses on textural and pattern similarity by calculating the distance between \textbf{pixel-level}, not learned, features.
In the case of having $N_{trans}$ translated images and $N_{tgt}$ target domain images to compare such that $N_{trans}\neq N_{tgt}$, we randomly sample $N=\mathrm{min}(N_{trans}, N_{tgt})$ images from each set to compute the distance metrics. We also note that FID may be inaccurate for our relatively small datasets ($N<2048$, where the feature covariance matrix is not full-rank); noting this, we report it as ``FID*''.

\section{Results and Discussion}
\begin{wrapfigure}[14]{htbp}{0.35\textwidth}
  \includegraphics[width=0.35\textwidth]{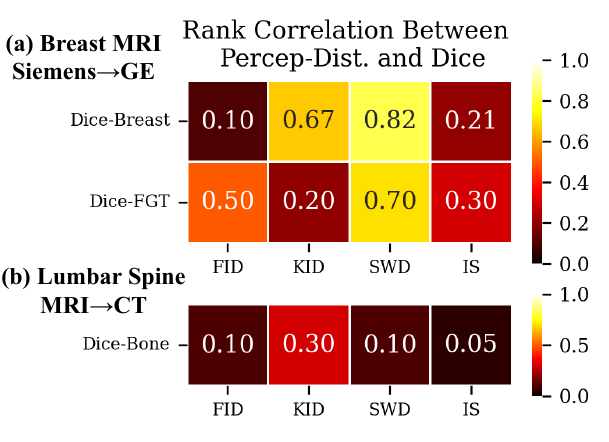}
  \caption{Absolute correlation of perceptual metrics with segmentation metrics.}
  \label{fig:corr}
\end{wrapfigure}

In Table \ref{tab:results} we show the performance of all translation models according to all metrics. For the perceptual metrics, a lower value is better for distances (FID, KID, SWD), and a higher one is better for IS. We also show example translated images in Fig. \ref{fig:imgs}, and the correlations of perceptual metrics with segmentation metrics are in Fig. \ref{fig:corr}.

\begin{table}[htbp]
\centering
\scriptsize
\begin{tabular}{l|cc|cccc||c|cccc}
\multicolumn{7}{c||}{\textbf{Breast MRI Siemens$\rightarrow$GE Translation}} & \multicolumn{5}{c}{\textbf{Lumbar Spine MRI$\rightarrow$CT Translation}}  \\
\toprule
\multicolumn{1}{c}{} & \multicolumn{2}{c}{Dice ($\uparrow$)} & \multicolumn{4}{c||}{Perceptual Metrics} & \multicolumn{1}{c}{Dice ($\uparrow$)} & \multicolumn{4}{c}{Perceptual Metrics}  \\
\cmidrule(lr){2-3}\cmidrule(lr){4-7}\cmidrule(lr){8-8}\cmidrule(lr){9-12}
\textbf{Method} &  Breast & FGT & FID* & KID  & SWD  & IS  & Bone & FID*  & KID  & SWD  & IS  \\
\midrule
\rowcolor{gray!30}\textit{None} & 0.927 & \underline{0.651} & 144 & 0.069 & 705 & 2.58 & 0.007 & 323 & 0.300 & 1553 & \textbf{2.93}\\
CycleGAN & \underline{0.934} & 0.529 & \textbf{107} & \textbf{0.049} & \underline{556} & 2.73 & \underline{0.229} & \underline{210} & \textbf{0.161} & \underline{960} & \underline{2.29} \\
MaskGAN & 0.865 & 0.277 & \underline{118} & 0.089 & 1037 & \textbf{3.00} & 0.158 & 248 & 0.217 & 1114 & 2.22 \\
UNSB & \underline{0.934} & 0.646 & 156 & 0.079 & 756 & 2.46 & 0.138 & \textbf{208} & \underline{0.172} & \textbf{932} & 2.14 \\
\hdashline  \vspace{-1em}\\
SPADE$^\dagger$ & \textbf{0.950} & \textbf{0.707} & 119 & \underline{0.067} & \textbf{500} & \underline{2.91} & \textbf{0.942} & 251 & 0.242 & 1359 & \underline{2.29} \\
\bottomrule
\end{tabular}
\caption{Quantitative results for both translation tasks. Best and runner-up models are shown in bold and underlined according to each metric, respectively.}
\label{tab:results}
\end{table}

Overall, \textbf{perceptual metrics do not consistently align with common segmentation metrics for medical image translation.} No single perceptual metric reliably correlates with segmentation metrics for both breast MRI and MRI-to-CT translation. Using a perceptual metric for model selection will depend highly on the choice of metric, with the commonly used FID being especially inconsistent.
Therefore, we advise caution in using FID for evaluating medical image translation.

SWD shows a better correlation than the learned feature metrics (FID, KID, IS) for the subtle intra-modality breast MRI translation. However, SWD fails for the more complex inter-modality translation of MRI-to-CT, likely due to its focus on pixel-level changes which are insufficient for capturing larger visual differences.

Given that perceptual metrics are designed for assessing image realism rather than object preservation during translation, their limited correlation with segmentation metrics is understandable. Nonetheless, this indicates that perceptual metrics may not be fully suitable for medical image translation. A broader evaluation approach and research into more universally applicable metrics are needed in this field.

% \suggest{Methods such as unconditional DDPM and UNSB maintain the consistency relying on that images with added noise will still retain structure from the original input. However, small but significant anatomical structure, for example FGT, may be lost in the noised version.}

\bibliography{main}

\end{document}